\documentstyle[twocolumn,aps,prd]{revtex}
\input{tcilatex}

\begin{document}
\title{Exploring the global topology of the universe}
\author{Helio V. Fagundes}
\address{Instituto de F\'{i}sica Te\'{o}rica, Universidade Estadual Paulista\\
S\~{a}o Paulo, SP 01405-900, Brazil\\
E-mail: helio@ift.unesp.br}
\maketitle

\begin{abstract}
In this talk work done by our group on cosmic topology\ is reviewed. It
ranges from early attempts to solve a famous controversy about quasars
through the multiplicity of images, to quantum cosmology in this context and
an application to QED renormalization.
\end{abstract}

\section{Introduction \ \ \ \ }

The preferred cosmological models for the description of the universe,
except for its very first instants, are those of
Friedmann-Lema\^{i}tre-Robertson-Walker (FLRW), popularly known as `Big
Bang' models. See, for example, Ohanian and Ruffini\cite{Ohanian}.

It is generally assumed that the global topology of cosmic spatial sections
is simply connected, that is, they are one of the spherical $(S^{3}),$
Euclidean $\left( E^{3}\right) $, or hyperbolic $(H^{3})$ spaces, which have
positive, null, and negative curvature, respectively.

But spaces $E^{3}$ and $H^{3}$ have infinite extension. This has led some
pioneers, like Einstein, Schwarzschild, and Friedmann, to consider the
question of the finite vs. infinite size of the universe, and its topology.

Beginning in 1971, with a paper by Ellis\cite{Ellis}, research on the
possibility of cosmic space being finite and its topology multiply
connected, has been done more systematically, if slowly at first.

Today there is a reasonable number of researchers in this area (my guess is
about a hundred worldwide), including here in Brazil.

Since this area is still little known, in this talk I will present a summary
of work done by myself and a few collaborators at IFT/UNESP in S\~{a}o
Paulo. The sections that follow list a number of published papers, which are
representative of the effect of nontrivial cosmic \ topology on the topics:
\ section II, formal theoretical works; III, the quasar redshift
controversy; IV, fitting models to data; V, cosmic crystallography; VI,
quantum field theory and quantum cosmology; and VII, the cosmic microwave
background.

Note that most of this research was done before 1998, so we could with
impunity assume a null cosmological constant.

\section{Formal theoretical works}

\subsection{A cosmological model with compact space sections and low mass
density\protect\cite{CMC}.}

In this model the adopted spacetime topology was $M^{4}=R\times \Sigma $,
where $R$ is the time axis and $\Sigma $ is the space section. This form of
spacetime holds for all works cited below (with different $\Sigma $'s, of
course).

Here $\Sigma =T_{g}\times S^{1},$ where $T_{g}$ is the genus-$g$ surface and 
$S^{1}$ is the circle. The metric is 
\[
ds^{2}=a^{2}(\eta )(d\eta ^{2}-d\rho ^{2}-\sinh ^{2}\rho \ d\varphi
^{2})-b^{2}(\eta )d\zeta ^{2}\ , 
\]
where 
\begin{eqnarray*}
ct(\eta ) &=&a_{\ast }(\sinh \eta -\eta )\ , \\
a(\eta ) &=&a_{\ast }(\cosh \eta -1)\ , \\
b(\eta ) &=&3a_{\ast }[\eta \coth (\eta /2)-2]\ ,
\end{eqnarray*}
with $a_{\ast }$ = constant. This metric had been studied by Kantowski and
Sachs \cite{KS}, and is of type III in the classification of Bianchi\cite
{Bianchi}.

At the time I was unaware of the existence of three-dimensional closed
hyperbolic manifolds, and this was the best substitute I could find for a
closed Friedmann model of undercritical density.

\subsection{Compactification of Friedmann's hyperbolic model\protect\cite
{CFH}}

Now $\Sigma $ is a {\it closed hyperbolic manifold} constructed by Best \cite
{Best}. I did not go into the details of image formation, but found that the
space of images was the covering space $H^{3}$, and that homogeneity and
isotropy should be reinterpreted in terms of the distribution of multiple
images in the covering space. I also found that we should have the relative
density of matter $\Omega _{0}<0.964$ in order that a repeated pattern of
images could be observed.

\subsection{Relativisic cosmologies with closed, locally homogeneous spatial
sections\protect\cite{RCC}}

Here I established a correspondence between the geometric classifications of
Bianchi and Kantowski-Sachs (BKS) on one side, and Thurston's on the other,
according to the following table:

\begin{center}
\begin{tabular}{|c|c|}
\hline\hline
{\bf Thurston type} & {\bf BKS type} \\ \hline
T1 & BIX \\ 
T2 & BI, BVII(0) \\ 
T3 & BV, BVII(A%
%TCIMACRO{\TEXTsymbol{>}}%
%BeginExpansion
\mbox{$>$}%
%EndExpansion
0) \\ 
T4 & KS \\ 
T5 & BVI(1) \\ 
T6 & BVIII \\ 
T7 & BII \\ 
T8 & BVI(0) \\ \hline\hline
\end{tabular}
\end{center}

It was also found that Bianchi types $IV$ and $VI(A),\ 0<A<1,$ cannot be
compactified - i. e., there are not any closed, locally homogeneous
manifolds of these Bianchi types.

\subsection{Closed spaces in cosmology\protect\cite{CSC}}

This paper improves on the previous one, and discusses the question of local
vs. global homogeneity of the constant curvature models.

\subsection{Numerical study of a perturbed Einstein-de Sitter cosmological
model\protect\cite{NSP}}

Here $\Sigma $ is the 3-torus $T^{3}$. Our purpose was to find how an
inhomogeneity of the matter density, written as $\rho (x,\tau ),\ \tau =1-t,$
affects the metric, which was assumed to have the form 
\[
ds^{2}=c^{2}d\tau ^{2}-a^{2}(x,\tau )dx^{2}-b^{2}(x,\tau )(dy^{2}+dz^{2})\ , 
\]
where $a^{2}(x,\tau )$, $b^{2}(x,\tau )$ reduce to $%
a_{EdS}(t)=(t/t_{0})^{2/3}$ if we remove the perturbation. The result was
that, for a perturbation of $\rho $ of about 80\%, the metric fluctuated at
most 5\%, in agreement with a heuristic estimate of Barrow\cite{Barrow}

\section{The quasar redshift controversy}

\subsection{Quasar-galaxy associations with discordant redshifts as a
topological effect. II. A closed hyperbolic model\protect\cite{QGAII}}

$\Sigma $ is one of Best's\cite{Best} manifolds with a regular icosahedron
as {\it fundamental polyhedron }(FP). The metric is the same as in
Friedmann's open model.

The search for images in conjunction to simulate the controversial pairs was
done by computer, by scanning all directions that met the observer. The
obtained pairs were not realistic: e. g., redshifts 0.12 for the galaxy and
4.3 for the quasar.

\subsection{Smallest universe of negative curvature\protect\cite{SUN}}

$\Sigma $ is the smallest known CHM, with normalized volume 0.9427... and 18
faces, discovered independently by Weeks\cite{Weeks} and by Matveev and
Fomenko\cite{Matveev}. It was assumed $\Omega _{0}=0.1.$ Here I got better
results for the quasar-galaxy associations than in the previous case: many
of the simulated pairs had realistic redshifts, like 0.002 for the galaxy
and 1.31 for the quasar or quasi-stellar object (QSO).

However, comparison with Burbidge et al.'s\cite{Burbidge} catalog of
associations, for example, would still not be reliable: the model parameters
are arbitrary, only 31 sources were used in the simulation, and of course
many conjunctions are line-of-sight concidences.

The model is also interesting on account of the small volume: multiple
images would allow astronomers to see objects at different epochs of their
evolution; and it has a larger probability than a larger model for the
spontaneous creation of the universe.

\section{Fitting models to data}

\subsection{A search for QSO's to fit a cosmological model with flat, closed
spatial sections\protect\cite{SQF}}

Based on a catalog with about 1500 quasars, we looked for pairs equidistant
from Earth, and along three orthogonal axis. $\Sigma $ was the 3-torus, with
a rectangular parallepiped as FP, whose sides were fitted to 3591, 2966, and
2792 Mpc.

\subsection{A suggestion on the pair of QSO triplets
1130+106\{B,A,C\},\{X,Y,Z\}\protect\cite{SPQ}}

The quasars in the title are two aligned triplets discovered by Arp and
Hazard\cite{AH}, with similar corresponding redshifts, which are \{2.1,
0.54, 1.6\} for the first triplet, and \{2.1, 0.51, 1.7\} for the second
one. They were adjusted to an Einstein-de Sitter model with a cube of side
387 Mpc as FP, which is an unrealistically small size.

\section{Cosmic crystallography}

\subsection{On closed Einstein-de Sitter universes\protect\cite{CES}}

The method of cosmic crystallography (CC) introduced by Lachi\`{e}ze-Rey et
al.\cite{LaLuUz} was applied to a few models, all with Einstein-de Sitter
metric but varying topologies. The idea was to verify that the CC method
yields observable (in principle) results, even if the FP's sides are of the
order of the{\it \ horizon}'s radius $R_{H}=2c/H_{0}$, where $H_{0}$ is
Hubble's constant. Good results were obtained with the FP's sides equal $%
0.7H_{0},$ and not so good ones with sides equal $1.2H_{0}$

\subsection{Cosmic crystallography in a compact hyperbolic universe 
\protect\cite{CCC}}

$\Sigma $ and FP are the same as in\cite{QGAII}. We applied the CC method to
this geometrically inhomogeneous model and did not obtain any sharp peaks -
as expected from the independent works of Lehoucq et al.\cite{LeLuUz} and
Gomero et al.\cite{GTRB}.

But we did get a smaller fluctuation that is not present in the control open
cosmology. We are at present tentatively attributing these small
fluctuations to {\it type II pairs}, as defined{\it \ }in \cite{LeLuUz}.

\section{Quantum field theory and quantum cosmology}

\subsection{A cosmic lattice as the substratum of quantum fields\protect\cite
{CLS}}

This work is a lowest-order attempt to avoid the infinities of quantum
electrodynamics renormalization. Both configuration and momentum spaces are
assumed to have the topology of a 3-torus $T^{3},$ with cubes of sides $%
a=c/H_{0}=8000$ Mpc = $2.7\times 10^{28}$ cm, and $P=2\pi \hbar /a$ $\approx
10^{20}$ Gev/$c,$ respectively.

Lorentz invariance is abandoned for very large, presently inaccessible
energies.

For charge and mass renormalization at the one-loop order, I got $%
Z_{3}=0.925 $, $e=\sqrt{Z_{3}}e_{0}=0.962e_{0},$ $m=1.185m_{0},$ and $%
Z_{2}=1.160$, in the notation of Itzykson and Zuber\cite{IZ}.

\subsection{Casimir energy in a small volume, multiply connected, static
hyperbolic preinflationary universe\protect\cite{Casimir}}

This work was orally communicated at this Meeting, and appears elsewhere in
these {\it Proceedings}. \ 

\subsection{Birth of a closed universe of negative spatial curvature 
\protect\cite{BCU}; On the birth of a closed hyperbolic universe\protect\cite
{BCH}}

$\Sigma $ is the first the lens space $L(50,1)$, then Weeks manifold as in 
\cite{SUN}.

Starting with the spontaneous creation, from a spherical orbifold, of a de
Sitter universe with the topology of a lens space, we proceeded as De
Lorenci et al.\cite{LMPS}, to obtain a quantum topology change into a de
Sitter universe with Weeks manifold as space section. After inflation it
becomes a closed hyperbolic universe, with the metric of an open FLRW
universe.

\section{The cosmic microwave background radiation}

\subsection{The quadrupole component of the relic radiation in a
quasi-hyperbolic cosmological model\protect\cite{QCR}}

This was an application of the model in\cite{CMC} to a quadrupole moment of
the cosmic microwave background reported by Fabbri and Melchiorri\cite{FM}.
Comparing the obtained relation 
\[
1+Z=\frac{a(\eta _{observ})}{a(\eta _{emission})}(1-\varepsilon \cos
^{2}\theta _{observ}) 
\]
with the result of\cite{FM}, I obtained $\varepsilon =4Q/2.7$K $=0.0013,$ $%
\Omega _{0}=1-3\varepsilon /2=0.998.$

\subsection{Fitting hyperbolic universes to Cay\'{o}n-Smoot spots in COBE
maps\protect\cite{fhu}; Sources of CMB spots in closed hyperbolic universes 
\protect\cite{SCMB}}

Cay\'{o}n and Smoot\cite{CS} identified several spots in NASA's satellite
COBE's maps of the cosmic microwave background\ as physical (rather than
noise), hence as small fluctuations of the matter density.

In these papers I simulated the position of six cold and eight hot CS spots
in closed hyperbolic manifolds, and interpreted them as having evolved into
today's galaxy superclusters and galaxy voids, respectively.

Present catalogs\cite{Einasto} only list superclusters up to $Z=0.12$, which
is completely inadequate for a fit of the simulated models: typically, in
one of the latter the 14 redshifts are in the range $0.361$ to $1.370$%
.\medskip

\smallskip I thank the Brazilian agency CNPq for partial financial support.

\end{document}